\documentclass[a4paper,aps,nofootinbib,showpacs,12pt]{revtex4}

\usepackage{graphicx}
\usepackage{amsfonts}
\usepackage{amsmath}
\usepackage{amsbsy}
\usepackage{longtable}
\usepackage{bm}
\usepackage{hyperref}
\usepackage{float}

\begin{document}
\title{A relativistic dissipative hydrodynamic description for systems
including particle number changing processes}

\author{Andrej El$^1$\footnote{el@th.physik.uni-frankfurt.de}, Azwinndini
Muronga$^{2}$\footnote{Azwinndini.Muronga@uct.ac.za}, Zhe
Xu$^{1,3}$\footnote{xu@th.physik.uni-frankfurt.de}, Carsten
Greiner$^{1}$\footnote{Carsten.Greiner@th.physik.uni-frankfurt.de}}

\affiliation{$^1$ Institut f\"{u}r Theoretische Physik, 
Goethe-Universit\"{a}t Frankfurt,
Max-von-Laue Strasse 1, D-60438, Frankfurt am Main, Germany}
\affiliation{$^2$ Institute for Theoretical Physics and Astrophysics, 
Department of Physics, University of Cape Town,
Rondebosch 7701, South Africa;
 UCT-CERN Research Centre, Department of Physics, University of Cape
Town, Rondebosch 7701, South Africa}
\affiliation{$^3$ Frankfurt Institute for Advanced Studies,
Ruth-Moufang-Strasse 1, D-60438, Frankfurt am Main, Germany}

\begin{abstract}
Relativistic dissipative hydrodynamic equations are extended by taking
into account particle number changing processes in a gluon system, which
expands in one dimension boost-invariantly.
Chemical equilibration is treated by a rate equation for the particle
number density based on Boltzmann equation and Grad's ansatz for the 
off-equilibrium particle phase space distribution.
We find that not only the particle production, but also the temperature
and the momentum spectra of the gluon system, obtained from the hydrodynamic
calculations, are sensitive to the rates of particle number changing processes.
Comparisons of the hydrodynamic calculations with the transport ones
employing the parton cascade BAMPS show the inaccuracy of the rate equation
at large shear viscosity to entropy density ratio.
To improve the rate equation, the Grad's ansatz has to be modified beyond
the second moments in momentum.
\end{abstract}

\pacs{47.75.+f, 24.10.Lx, 24.10.Nz, 12.38.Mh, 25.75.-q}

\maketitle

\section{Introduction}
Relativistic dissipative hydrodynamics has established itself as 
an effective theory for investigations of phenomena in relativistic heavy-ion 
collisions owing to its success in describing the elliptic flow $v_2$ data
measured at the Relativistic Heavy Ion Collider (RHIC) at Brookhaven National
Laboratory (BNL)
\cite{Voloshin:2008dg,Romatschke:2007mq,Song:2007ux,Luzum:2008cw,Heinz:2009xj,Teaney:2009qa}. 
Most hydrodynamic approaches 
\cite{Romatschke:2007mq,Song:2007ux,Luzum:2008cw,Molnar:2008fv,Molnar:2009tx}
do assume an instantaneous chemical 
equilibrium of matter constituents. However, this requires an infinite large
transition rate of microscopic processes which drive the system towards the
chemical equilibrium and thus breaks the consistency of the physical
description for matter with a finite viscosity. Another extreme but
physically consistent case, as considered in Ref. \cite{HM09}, is to 
keep the number of the constituents constant by assuming that there are
no particle number changing processes. In this case deviation of
the one-particle phase space distribution from its chemical equilibrium 
form becomes larger and larger when the system expands. In real 
ultrarelativistic heavy-ion collisions bremsstrahlung processes of gluons 
and quarks, also gluon fusion to quark and antiquark, and the reverse
processes govern the chemical equilibration of gluons and 
quarks \cite{Biro93,Srivastava:1996qd,ER2000,Xu:2004mz,Gelis:2005pb}. 
Therefore, particle number changing processes with realistic rates should
be included in underlying hydrodynamic descriptions for quark-gluon plasma
(QGP).

Because the energy density $e \sim \lambda T^4$, where $\lambda$
is the fugacity and $T$ the temperature, for keeping the energy density
unchanged the temperature decreases by $20\%$ when the fugacity increases
from $0.5$ to $1$. Thus, an effect of chemical equilibration is expected 
to be present for the temperature, which is an important detail, for instance
for the dilepton
\cite{Srivastava:1996qd,Dusling:2008xj,Martinez:2008di,vanHees:2009vk} 
and photon \cite{Schenke:2006yp} yields in ultrarelativistic heavy-ion
collisions. Moreover, to what degree chemical equilibration of quarks and
gluons is achieved at the phase transition might also be crucial for 
modelling hadronization using recombination models \cite{Lin:2002rw}.

In this work we develop a dissipative hydrodynamic approach including
particle number changing processes via a rate equation to investigate
the bulk properties of the QGP at RHIC. As a first step we simplify the 
expansion dynamics at RHIC by assuming a one-dimensional expansion in
the beam axis with Bjorken boost invariance. The QGP is considered to
possess a constant $\eta/s$ value during its evolution. Different from
the previous studies of chemical equilibration of the QGP basing on 
either ideal \cite{Biro93, Srivastava:1996qd} or
first-order \cite{Dutta:1999cn} hydrodynamic equations, we apply the
second-order Israel-Stewart \cite{HM09,El09} as well as the extended
third-order hydrodynamic equations \cite{3rdO} to describe the QGP
evolution. The hydrodynamic equations are now coupled with a rate
equation for the particle number density. The microscopic interactions,
which are responsible for the viscous behaviour and change of the
particle number, are simplified by elastic binary $2\to 2$ and
inelastic multiplication and annihilation $2\leftrightarrow 3$ processes
with isotropically distributed collision angles. Although the evolution
of multi-component systems as discussed in Ref. \cite{Monnai:2010qp}
is gaining an increasing attention, here we consider only a pure gluon
system for the sake of simplicity.

To examine the applicability of the new description we compare the
hydrodynamic solutions with those calculated from a parton cascade,
the Boltzmann approach of multiparton scatterings (BAMPS) \cite{Xu:2004mz},
in a similar manner as done in 
Refs. \cite{HM09,El09,3rdO,Bouras:2009nn,Denicol:2010xn}.
The $\eta/s$ is extracted from the BAMPS calculations using the procedure
introduced by us in Ref. \cite{El09}.

The paper is organised as follows. In Sec. \ref{hydro_eq} we introduce 
the dissipative hydrodynamic equations for a one-dimensional boost-invariant 
expanding system of gluons in presence of particle production and annihilation
processes. Solutions are compared with those without the inelastic processes
and those assuming chemical equilibrium. This quantitatively demonstrates the
effects of chemical equilibration on physical observables. 
Influences of the initial conditions on these observables is studied in Sec. \ref{init}. In Sec.
\ref{hydro_vs_BAMPS} we compare
the results of hydrodynamic calculations with those from BAMPS.
Conclusions are given in Sec. \ref{conclusions}.

\section{Viscous hydrodynamic equations and rate equation describing chemical equilibration}
\label{hydro_eq}
Derivations of dissipative hydrodynamic equations either from the 
entropy production principle or the moments method were reported in 
Refs. \cite{IS,M04,3rdO}. Ignoring the bulk viscosity and the heat 
conductivity, one obtains the evolution equation for the shear tensor
\begin{equation}
\underbrace{\dot{\pi}^{\alpha\beta}  =  -\frac{\pi^{\alpha\beta}}{\tau_\pi} +
\frac{\sigma^{\alpha\beta}}{\beta_2} - \pi^{\alpha\beta} \frac{T}{\beta_2}
\partial_\mu \left( \frac{\beta_2}{2T}u^\mu \right)}_{\text{IS}} +  
\underbrace{\alpha \frac{T}{\beta_2} \partial_\mu \left( \frac{\beta_2^2}{T} 
u^\mu \right)\pi^{\langle \alpha}_{\sigma}
\pi^{\sigma\beta\rangle}}_{\text{third-order}} \,,
\label{3rd_order_pimunu}
\end{equation}
where
\begin{equation}
\sigma^{\alpha\beta}=\nabla^{\langle\alpha}u^{\beta\rangle}=\left(\frac{1}{2}
(\Delta_{\mu}^{\alpha}\Delta_{\nu}^{\beta}+\Delta_{\mu}^{\beta}
\Delta_{\nu}^{\alpha})-\frac{1}{3}\Delta_{\mu\nu}\Delta^{\alpha\beta}
\right)\nabla^{\mu}u^{\nu}
\end{equation}
and $\Delta_{\alpha\beta}=g_{\alpha\beta}-u_\alpha u_\beta$ with
the metric $g_{\alpha\beta}={\rm diag} (1,-1,-1,-1)$. 
$u^\mu$ is the fluid velocity with $u_\mu u^\mu=1$.
$\tau_\pi=2\eta\beta_2$ is the relaxation time, where $\eta$ denotes 
the shear viscosity and $\beta_2=9/(4e)$.
The ``IS'' part is exactly the equation from the Israel-Stewart
theory, while the ``third-order'' part indicates a higher order correction
term derived in Ref. \cite{3rdO}. We obtained $\alpha=-8/9$.

We simplify the dynamical evolution of a QGP created in ultrarelativistic
heavy-ion collisions by a one-dimensional boost-invariant expansion of 
massless Boltzmann particles. For this case $u^\mu=(t,0,0,z)/\tau$ where
$\tau=\sqrt{t^2-z^2}$ is the proper time. With $\partial_\mu u^\mu=1/\tau$
and $u^\mu\partial_\mu=\frac{d}{d\tau}$ Eq. (\ref{3rd_order_pimunu}) reduces to
\begin{equation}
\dot {\bar \pi} = -\frac{{\bar \pi}}{\tau_\pi} +
\frac{8}{27}\frac{e}{\tau} -\frac{1}{2}\frac{{\bar \pi}}{\tau} +
\frac{1}{2}{\bar \pi}\frac{\dot T}{T} +
\frac{1}{2}{\bar \pi}\frac{\dot e}{e}+
\frac{3}{2}\frac{{\bar \pi}^2}{e\tau} +
\frac{3}{2}\frac{{\bar \pi}^2}{e}\frac{\dot T}{T} + 
\frac{3}{2}\frac{{\bar \pi}^2}{e}\frac{\dot e}{e} - 
4\frac{{\bar \pi}^2}{e\tau} \,,
\label{pi}
\end{equation}
where ${\bar \pi}=\pi^{33}=-2\pi^{11}=-2\pi^{22}$ is the only independent
component of $\pi^{\mu\nu}$ in the local rest frame. 
The dot denotes the derivative with respect to $\tau$.
Time evolution equation for the energy density $e$ follows from the
conservation of energy, $\partial_\mu T^{\mu 0}=0$, where $T^{\mu\nu}$ is
the energy-momentum tensor. For a one-dimensional boost-invariant system one
obtains \cite{3rdO, M04}
\begin{equation}
 \dot e = -\frac{4}{3}\frac{e}{\tau} + \frac{{\bar \pi}}{\tau} \,.
\label{e}
\end{equation}

Equation (\ref{pi}) corresponds to the Grad's ansatz \cite{3rdO} for an 
off-equilibrium distribution of particles in phase space
\begin{equation}
 f(x,p) = f_{eq}(x,p) \lambda \left[ 1 + \frac{3 {\bar \pi}}{8 e T^2}
\left( \frac{1}{2}p_T^2 - p_z^2 \right) \right] \,,
\label{f_offeq_1d}
\end{equation}
where $f_{eq}(x,p)=g e^{-E/T}$ is the equilibrium distribution. $g$ is the
degeneracy factor and $g=16$ for gluons. $\lambda$ denotes the fugacity
and quantifies the degree of the chemical equilibration. The distribution
(\ref{f_offeq_1d}) satisfies the matching of the particle number and energy
densities, $n = \lambda n_{eq} =  \lambda g T^3/\pi^2$ and 
$e = \lambda e_{eq} = 3 \lambda g T^4/\pi^2$, by which an off-equilibrium
state is matched to a fictitious equilibrium state \cite{M07}. The matching
conditions allow to define temperature for an off-equilibrium system.
Assuming chemical equilibrium, i.e., $\lambda=1$, one obtains 
\begin{equation}
T = \left( \frac{\pi^2 e}{3g} \right)^{1/4} \,.
\label{temp1}
\end{equation}
With this one can solve the coupled equations (\ref{pi}) and (\ref{e})
once the initial conditions are given. The assumption of chemical
equilibrium is made in most of the hydrodynamic calculations 
\cite{Romatschke:2007mq,Song:2007ux,Luzum:2008cw,Molnar:2008fv,Molnar:2009tx}.
However, an instantaneous and complete chemical equilibrium requires
an infinite transition rate of particle number changing processes, which
is not consistent with a finite shear viscosity. Without the assumption of
the chemical equilibrium, the temperature 
\begin{equation}
T=\frac{e}{3n}
\label{temp2}
\end{equation}
depends also on the particle number density. Thus, we need an additional
equation for the time evolution of $n$. 

We assume that the space-time evolution of  gluons obeys the Boltzmann equation
\begin{equation}
p^\mu \partial_\mu f(x,p)= C[f] \,,
\label{boltzmann}
\end{equation}
where $C[f]$ is the collision integral and contains microscopic details
of particle interactions. Comparisons between the solutions of the 
hydrodynamic equations and those of the Boltzmann equation will be
shown in Sec. \ref{hydro_vs_BAMPS}. Here we use the Boltzmann equation
to derive an equation for the particle number density $n$.

Integration of Eq. (\ref{boltzmann}) over $d\Gamma=d^3 p/[(2\pi)^3 E]$ leads to
\begin{equation}
 \int d\Gamma p^\mu \partial_\mu f(x,p) =  \int d\Gamma C[f] \,.
\label{int_C}
\end{equation}
The left hand side can be rewritten as the derivative of the particle
number current, $\partial_\mu N^\mu$, which is equal to $\dot n + n/\tau$
for a one-dimensional boost-invariant expansion. The right hand side
presents the source of particle production and annihilation. Considering
two to three particles and vice versa as the only particle number changing
processes, one obtains the rate equation 
\cite{Biro93,Srivastava:1996qd,Dutta:1999cn,Muronga05}
\begin{equation}
 \dot n + \frac{n}{\tau} = \frac{1}{2} n R_{23} - \frac{1}{3} n 
R_{32}  \,,
\label{master_eq}
\end{equation}
where $R_{23}$ and $R_{32}$ denote the collision rates per particle
for inelastic $2\to 3$ and $3\to 2$ processes. The factor $1/2$ and $1/3$
indicate that colliding particles are identical. The rates are given 
by \cite{Xu:2007aa,Xu:2010cq}
\begin{equation}
R_{23} = n \langle v_{\rm rel} \sigma_{23} \rangle_2 \,, \quad
R_{32} = \frac{1}{2} n^2 
\left \langle \frac{I_{32}}{8E_1E_2E_3} \right \rangle_3\,,
\end{equation}
where the averages are defined as
\begin{eqnarray}
\langle {\cal Q} \rangle_2 &=& \frac{1}{n^2} \int \frac{d^3p_1}{(2\pi)^3}
\frac{d^3p_2}{(2\pi)^3} \
f(x,p_1) f(x,p_2)\ {\cal Q} \,,\\ 
\langle {\cal Q} \rangle_3 &=& \frac{1}{n^3} \int \frac{d^3p_1}{(2\pi)^3}
\frac{d^3p_2}{(2\pi)^3} \frac{d^3p_3}{(2\pi)^3} \
f(x,p_1) f(x,p_2) f(x,p_3) \ {\cal Q} \,.
\end{eqnarray}
$v_{rel}=(p_1+p_2)^2/(2E_1E_2)$ is the relative velocity of two colliding
particles with four momenta $p_1$ and $p_2$. The general definitions
of the cross section $\sigma_{23}$ and $I_{32}$ can be found in
Refs. \cite{Xu:2004mz,Xu:2010cq}. For interactions with isotropic 
distributions of collision angles $I_{32}$ is related to $\sigma_{23}$ 
via $I_{32}=192\pi^2 \sigma_{23}$ \cite{Xu:2004mz}. Using the approximate
distribution (\ref{f_offeq_1d}) we obtain
\begin{equation}
\frac{1}{2} n R_{23} - \frac{1}{3} n R_{32} =
\frac{1}{2} n^2 (1-\lambda) \sigma_{23} \,,
\label{approx}
\end{equation}
which leads to
\begin{equation}
 \dot n + \frac{n}{\tau} = \frac{1}{2} n^2 (1-\lambda) \sigma_{23} \,.
\label{n}
\end{equation}
Equation (\ref{n}) is similar to the rate equations derived in 
Ref. \cite{Biro93}.

The hydrodynamic equations (\ref{pi}), (\ref{e}), and (\ref{n}) 
describe kinetic and chemical equilibration of a gluon matter. 
Whereas the value of $\eta$ does not depend on the types of microscopic
interactions \cite{Xu:2007aa}, the net particle production, which
explicitly affects the shear pressure ${\bar \pi}$ and $n$, is of course
strongly dependent on the strength of inelastic collisions.

The standard viscous hydrodynamic approaches like the ones employed in
Refs. \cite{Song:2007ux,Luzum:2008cw,Molnar:2008fv} are up to second order in
gradients and dissipative quantities. In the one-dimensional case
studied here, the relevant gradient is the expansion scalar 
$\partial_\mu u^\mu = 1/\tau$ and the relevant dissipative quantity
is the shear pressure ${\bar \pi}$. Both $\tau_\pi/\tau$ and ${\bar \pi}/e$
must be small to ensure the validity of the hydrodynamic approach.
In our notation a term of $q$-th order has the form 
$(\tau_\pi/\tau)^{q_1}({\bar \pi}/e)^{q_2}$
with $q_1+q_2=q$. To make a proper order counting for Eq. (\ref{pi})
we rewrite $\dot T/T=\dot e/e-\dot n/n$ according to Eq. (\ref{temp2})
and insert Eqs. (\ref{e}) and (\ref{n}) for $\dot e$ and $\dot n$ 
into (\ref{pi}). We then multiply Eq. (\ref{pi}) by $\tau_\pi/e$.
Keeping terms up to second order we obtain
\begin{equation}
\dot {\bar \pi} = -\frac{{\bar \pi}}{\tau_\pi} - 
\frac{4}{3}\frac{{\bar \pi}}{\tau} +
\frac{8}{27}\frac{e}{\tau} -
\frac{1}{4}{\bar \pi} n (1-\lambda)\sigma_{23} \,.
\label{pi_2ndO}
\end{equation}
We refer to Eq. (\ref{pi_2ndO}) as the Israel-Stewart (IS) equation.
A similar equation, used in Refs. \cite{HM09,Muronga05} assuming particle number 
conservation, i.e., $\sigma_{23}=0$, contains the term ${\bar \pi}^2/(e\tau)$
which is neglected here, because this term times $\tau_\pi/e$ is of 
third order. Keeping terms up to third order we obtain
\begin{equation}
\dot {\bar \pi} = -\frac{{\bar \pi}}{\tau_\pi} - 
\frac{4}{3}\frac{{\bar \pi}}{\tau} +
\frac{8}{27}\frac{e}{\tau} - 3 \frac{{\bar \pi}^2}{e\tau} -
\frac{1}{4}\left ( 1 + 3\frac{{\bar \pi}}{e} \right )
{\bar \pi} n (1-\lambda)\sigma_{23} \,.
\label{pi_3rdO}
\end{equation}
We refer to Eq. (\ref{pi_3rdO}) as the ``third-order'' equation.

In the following we solve the IS equation (\ref{pi_2ndO}) as well as 
the third-order equation (\ref{pi_3rdO}) coupled with
Eqs. (\ref{e}) and (\ref{n}). We consider a gluon system with 
a constant $\eta/s=0.35$, which is a rough estimate of the upper bound of
the $\eta/s$ value found at RHIC \cite{Luzum:2008cw}. The entropy density
$s$ is taken at kinetic equilibrium, i.e., at ${\bar \pi}=0$
\begin{equation}
s = 4n-n\ln\lambda \,.
\label{s}
\end{equation}
The fugacity is related to $e$ and $n$ via
$\lambda=n/n_{eq}=(\pi^2/g)nT^{-3}=(\pi^2/g)n(e/3n)^{-3}$.
Since $\eta \sim e \lambda_{mfp}$, where $\lambda_{mfp}$ is the gluon
mean free path, we have $\eta/s \sim (T/s)(n\lambda_{mfp})$. With the choice
\begin{equation}
\sigma_{23} = \frac{g}{\pi^2} \frac{T}{s}=
\frac{1}{\lambda (4-\ln\lambda) T^2} 
\label{sigma23}
\end{equation}
and the same cross section of binary collisions $\sigma_{22}=\sigma_{23}$,
a constant $\eta/s=0.35$ can be obtained in kinetic transport calculations,
as will be shown in Fig.\ref{fig:etaovers} in Sec. \ref{hydro_vs_BAMPS}.

To demonstrate effects of the chemical equilibration on observables
we compare results of calculations with non-vanishing $\sigma_{23}$ to those
with $\sigma_{23}=0$ (particle number conservation) and with $\lambda=1$
(instantaneous chemical equilibrium). In the case of $\sigma_{23}=0$
Eq. (\ref{n}) gives $\dot n=-n/\tau$ and leads to 
$n(\tau)=n(\tau_0) \tau_0/\tau$. For $\lambda=1$ we have
$n=n_{eq}=(3g/\pi^2)^{1/4}e^{3/4}/3$, which is the solution of Eq. (\ref{n})
only if ${\bar \pi}=0$, i.e., $\eta=0$. This corresponds to infinite 
transition rates of particle number changing processes, which keep the 
system in chemical equilibrium. Here we also see that the setup of
$\lambda=1$ and a non-vanishing shear viscosity $\eta$ is physically
inconsistent.

\begin{figure}[ht]
\centering
\includegraphics[width=14cm]{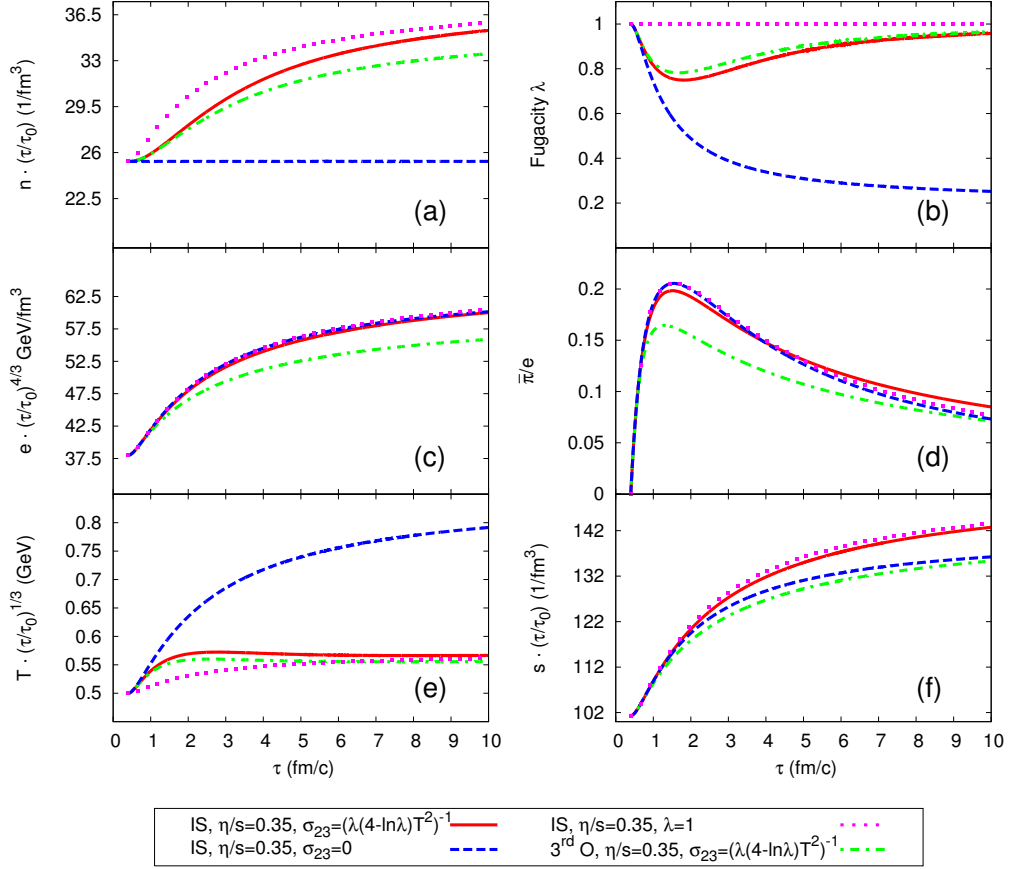}
\caption{(Color online) Time evolution of 
(a) the rescaled particle number density $n\cdot(\tau/\tau_0)$, 
(b) the fugacity $\lambda$, 
(c) the rescaled energy density $e\cdot(\tau/\tau_0)^{4/3}$,
(d) the ratio ${\bar \pi}/e$,
(e) the rescaled temperature $T \cdot(\tau/\tau_0)^{1/3}$ and 
(f) the rescaled entropy density $s \cdot(\tau/\tau_0)$.
The cases $\sigma_{23}=0$ (dashed curves) and $\lambda=1$ (dotted curves)
are calculated using the IS equations. The standard case with 
Eq. (\ref{sigma23}) is calculated using both the IS (solid curves) 
and the third-order equations (dash-dotted curves).
}
\label{fig:comp}
\end{figure}

Figure \ref{fig:comp} shows the results for a thermal initial condition
at $\tau_0 = 0.4 $ fm/c with $T(\tau_0) = 0.5 $ GeV and $\lambda(\tau_0)=1$.
To see the effects of the finite shear viscosity and the particle number
changing processes, we rescale the particle number density, the temperature,
the energy and the entropy density by their time evolution in an ideal fluid.
We first compare the results for non-vanishing $\sigma_{23}$ (solid curves), 
$\sigma_{23}=0$ (dashed curves) and $\lambda=1$ (dotted curves) using 
the IS equations. With the particle number conservation ($\sigma_{23}=0$)
$n\cdot \tau/\tau_0 = n(\tau_0)$ is constant. The non-vanishing $\sigma_{23}$
leads to a net increase of particle number, because the finite shear 
viscosity brings the system away from chemical equilibrium and the system
becomes undersaturated ($\lambda < 1$). This is clearly 
demonstrated in the time evolution of the fugacity in Fig. \ref{fig:comp} (b).
It follows from Eq. (\ref{n}) that the larger the value of $\sigma_{23}$,
the larger is the particle number increase and thus the faster is
the restoration of the chemical equilibrium. In the limit of instantaneous
restoration ($\lambda=1$), corresponding to $\sigma_{23}\to \infty$,
a maximum of particle productions is achieved, as shown by the dotted
curve in Fig. \ref{fig:comp} (a).

Due to the viscous effect, the decrease of the energy density is slower
than in the ideal fluid. Three curves in Fig. \ref{fig:comp} (c) differ
only marginally. Same is observed for the ${\bar \pi}/e$ ratio in
Fig. \ref{fig:comp} (d). Time evolution of the energy density and the shear
pressure depend mainly on the viscosity and not much on the details of 
microscopic interactions. Formally, the weak dependence of $e$ and 
${\bar \pi}$ on $\sigma_{23}$ or $\lambda$ is explained by the fact that 
the corresponding term in Eq. (\ref{pi}) is proportional to $\dot T/T$,
which has a logarithmic dependence on the temperature.

Because of $T=e/(3n)$ the rates of the particle number changing processes
influence the value of the temperature. The larger the rates, which are 
infinity in the case of the instantaneous equilibration ($\lambda=1$),
the smaller is the temperature, as shown in Fig. \ref{fig:comp}(e).
Between the two limits, $\sigma_{23}=0$ and $\lambda=1$, there is a $50\%$
difference at the final time $\tau=10 $ fm/c.

Due to the matching $n=\lambda n_{eq}$ and $e=\lambda e_{eq}$ we
obtain $n \sim e^{3/4}\lambda^{1/4}$ and thus 
$s \sim e^{3/4}\lambda^{1/4}(4-\ln \lambda)$ according to Eq. (\ref{s}).
Because the energy density is rather insensitive to the evolution of 
fugacity $\lambda$ [Fig. \ref{fig:comp} (c)] and the function 
$\lambda^{1/4}(4-\ln \lambda)$ is almost constant for not very small 
$\lambda$, the entropy density weakly depends on the fugacity, see 
Fig. \ref{fig:comp} (f). A larger effect can be observed at late times 
in the case of $\sigma_{23}=0$, because at that time the fugacity is 
quite small if particle number is conserved.

The third-order correction terms to the IS equations reduce the entropy
production. This is the main finding of Ref. \cite{3rdO}.
In Fig. \ref{fig:comp} (f) we observe that the entropy density calculated
from the third-order equations (dash-dotted curve) is considerably smaller 
than that from the IS equations (solid curve). Thus, in oder to obtain the
same $\eta/s$, a smaller shear viscosity $\eta$ is needed, which leads to
smaller energy density and shear pressure, compared with the results of the
IS equations. However, the interplay between these quantities and $n$, $T$,
and $\lambda$ appears to be non-trivial.

\begin{figure}[htb]
\centering
\includegraphics[width=12cm]{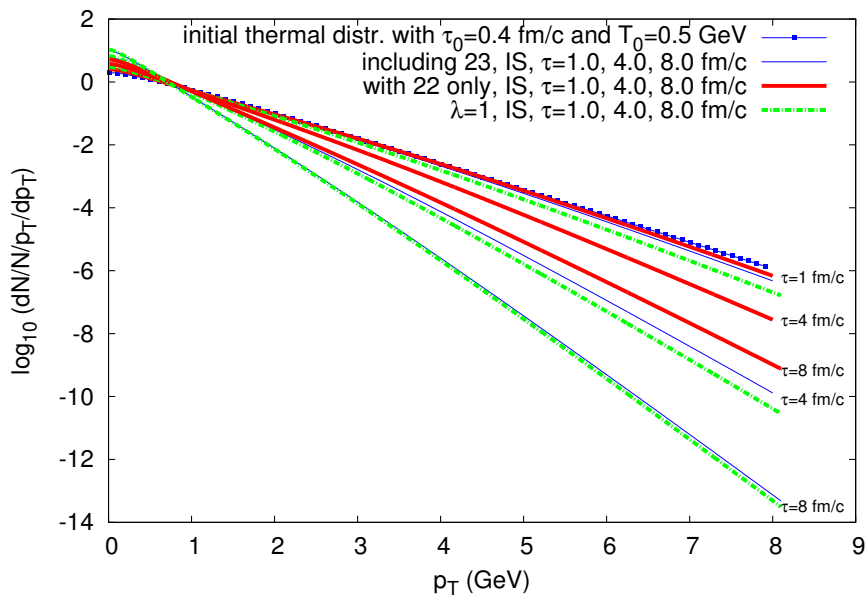}
\caption{(Color online) Transverse momentum spectra according to
the distribution (\ref{f_offeq_1d}). Depicted are the initial spectrum 
at $\tau_0=0.4$ fm/c and spectra at $\tau=1$ fm/c, $4$ fm/c and $8$ fm/c.
}
\label{fig:hydro_spectra}
\end{figure}

Since the rates of the particle number changing processes influence
the temperature, they should as well influence the particle momentum
spectrum, which is more closely related to experimental data.
Using the distribution (\ref{f_offeq_1d}) we calculate the particle
transverse momentum spectra at different times and demonstrate them
in Fig. \ref{fig:hydro_spectra}. Values of $e$, ${\bar \pi}$, $T$, and
$\lambda$ are taken from solutions of the IS equations in the three cases
shown in Fig. \ref{fig:comp}. According to Eq. (\ref{f_offeq_1d}) 
the spectrum slope is approximately proportional to the temperature 
regardless of the dissipative correction. Significant differences
in the spectra for the three cases at a late time, $\tau=4$ fm/c
or $\tau=8$ fm/c, reflect the differences in temperature shown in
Fig. \ref{fig:comp} (e).

\section{Influence of initial conditions}
\label{init}

In ultrarelativistic heavy-ion collisions initially produced quarks are
much less abundand than gluons. Chemical equilibration of quarks and gluons start
with different initial values of fugacity. The initial fugacity of
quarks is expected to be much smaller than one \cite{Biro93}. 
In this section we repeat calculations performed for Fig. \ref{fig:comp}
but with an initial fugacity $\lambda(\tau_0)=0.2$
instead of $1$. The initial condition is a kinetically equilibrated, but
chemically disequilibrated system. The initial energy density is
$e(\tau_0)=\lambda(\tau_0)e_{eq}(\tau_0)=(3g/\pi^2)\lambda(\tau_0)[T(\tau_0)]^4$
with $T(\tau_0)=0.5/[\lambda(\tau_0)]^{1/4}$ GeV. 
We calculate the IS equations, Eqs. (\ref{pi_2ndO}), (\ref{e}), and 
(\ref{n}), for $\eta/s=0.35$.

\begin{figure}[ht]
\centering
\includegraphics[width=14cm]{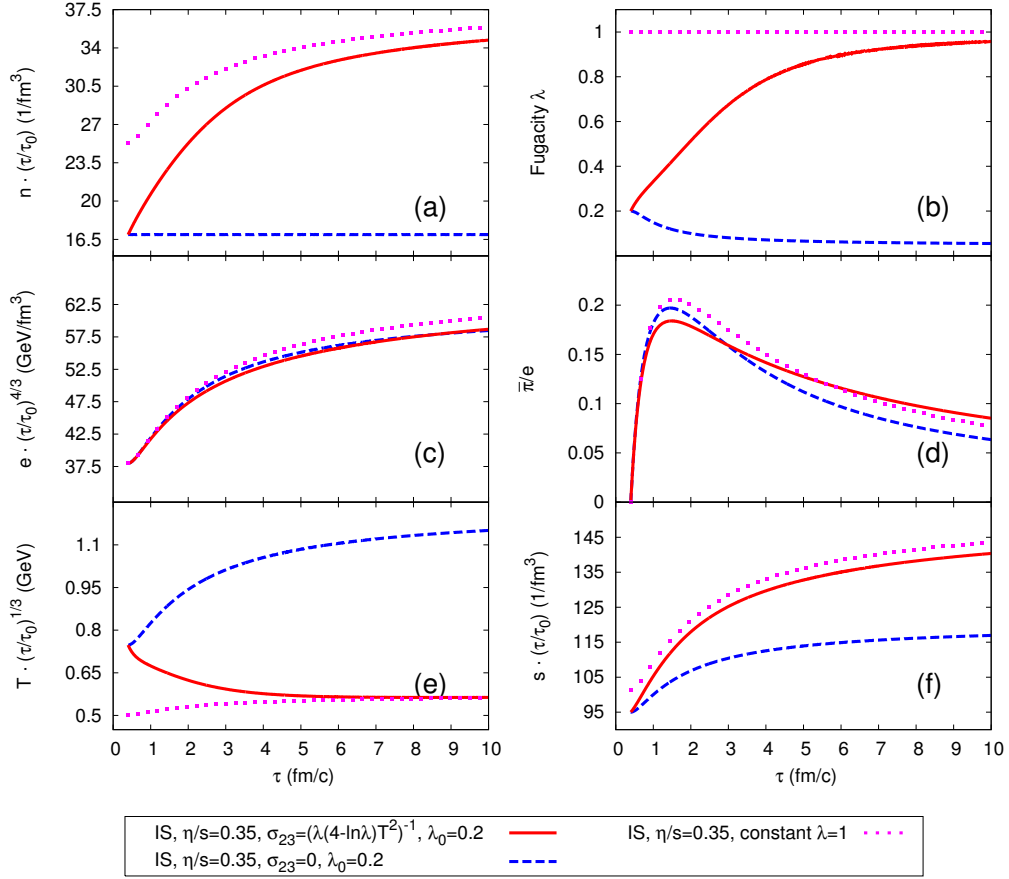}
\caption{(Color online) Same as Fig. \ref{fig:comp} using 
the IS equation (\ref{pi_2ndO}) with kinetically equilibrated and 
chemically disequilibrated initial condition,
$\lambda(\tau_0)=0.2$.}
\label{fig:comp2}
\end{figure}

The results are shown in Fig. \ref{fig:comp2} for three cases:
non-vanishing $\sigma_{23}$ [Eq. (\ref{sigma23})] by the solid curves,
$\sigma_{23}=0$ by the dashed curves, and $\lambda=1$ by the dotted curves.
In the case of $\lambda=1$ the initial 
temperature is  $0.5$ GeV; it is larger than $0.5$ GeV in case $\lambda_0=0.2$ to hold the same
initial energy
density. The initial particle number density and the entropy density
are larger in the $\lambda=1$ case compared with the values of the other two cases.

As already observed in Fig. \ref{fig:comp}, the energy density and the 
shear pressure, shown in Fig. \ref{fig:comp2} (c) and (d), are weakly
affected by the value of $\sigma_{23}$. On the contrary,
without particle number changing processes, 
$\sigma_{23}=0$, the fugacity [see Fig.\ref{fig:comp2} (b)] decreases and
the system goes far away from the chemical equilibrium, while the fugacity
increases to $1$ for large $\sigma_{23}$. Accordingly, pronounced
differences are seen in the particle number density [Fig. \ref{fig:comp2} (a)],
the temperature [Fig. \ref{fig:comp2} (e)], and the entropy density
[Fig. \ref{fig:comp2} (f)] between the non-vanishing 
$\sigma_{23}$ and $\sigma_{23}=0$ cases. We also observe that the results of 
$n$, $\lambda$ and $T$ in the non-vanishing $\sigma_{23}$ case relax to 
those of the $\lambda=1$ case. 

Comparisons in Fig. \ref{fig:comp2} underline the importance
of microscopic details in applying hydrodynamic approaches to describe
both chemical and kinetic equilibration of an initially undersaturated system.

\section{Comparisons between the hydrodynamic and transport calculations}
\label{hydro_vs_BAMPS}
Kinetic transport theory describes the space-time evolution of particles
by means of detailed treatment of microscopic processes. The dynamical 
behaviour of a particle system resembles the one described by viscous
hydrodynamics if the particle mean free path is much smaller
than macroscopic scales. Comparisons between hydrodynamic and transport 
calculations are always useful to examine the applicability limits of
viscous hydrodynamic theories
\cite{HM09,El09,3rdO,Bouras:2009nn,Denicol:2010xn}.

In this section we compare results from the new viscous hydrodynamic 
description including particle number changing processes presented in
Sec. \ref{hydro_eq} with results of transport calculations using the parton
cascade BAMPS \cite{Xu:2004mz}.
BAMPS simulates particle interactions, especially inelastic 
$2\leftrightarrow 3$ processes with full detailed balance, by using the 
stochastic interpretation of transition rates.

\begin{figure}[ht]
\begin{center}
  \includegraphics[width=12cm]{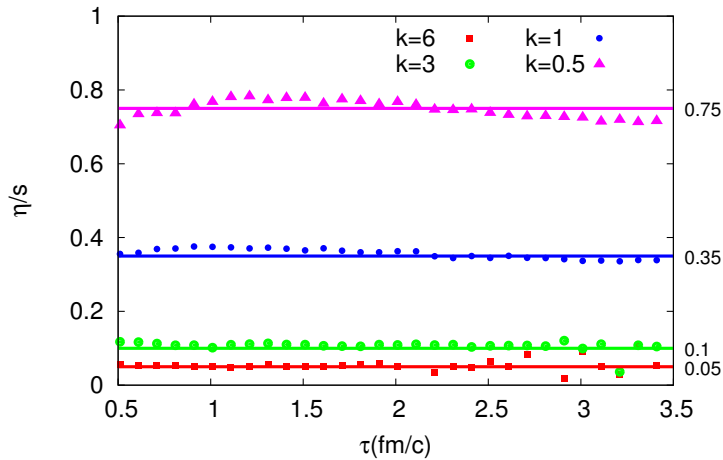}
\end{center}
\caption{(Color online) $\eta/s$ ratio extracted from BAMPS calculations
with time-dependent cross sections.}
\label{fig:etaovers}
\end{figure}

From the kinetic theory point of view, the physical origin of a finite
shear viscosity is the finite rates of microscopic collision processes.
The relation between the (transport) collision rate and the shear
viscosity coefficient was reported in Refs. \cite{XuPRL08,El09}. 
Here we use the formula derived in
\cite{El09}
\begin{equation}
 \eta=-\frac{\pi_{\mu\nu}\pi^{\mu\nu}}{2T C_0 \pi_{\mu\nu}P^{\mu\nu}}
\label{eta}
\end{equation}
with $P^{\mu\nu}=\int d\Gamma p^\mu p^\nu C[f]$ and 
$C_0=3/(8eT^2)$. The collision integral $C[f]$ contains details of
microscopic interactions. We consider elastic binary $2\to 2$ and 
inelastic $2\leftrightarrow 3$ processes with isotropic distributions
of collision angles. The cross sections are parametrized by 
\begin{equation}
 \sigma_{22}=\sigma_{23}=\frac{k}{\lambda (4-\ln\lambda) T^2} \,,
\label{cs}
\end{equation}
where $k$ is a parameter controlling the value of $\eta$. With this choice
one expects to obtain a constant $\eta/s$ as discussed around
Eq. (\ref{sigma23}). The transition probabilities of a $2\to 2$ or
$2\to 3$ process within a unit volume $\Delta v$ and a unit time step 
$\Delta t$ are same and given by 
$P_{23}=P_{22}=v_{rel} \sigma_{23} \Delta t/ \Delta v$.
The transition probability of a $3\to 2$ process is 
$P_{32}=(1/8E_1E_2E_3) I_{32} \Delta t/ \Delta v^2$
with $I_{32}=192\pi^2 \sigma_{23}$ \cite{Xu:2004mz}.
Figure \ref{fig:etaovers} shows the $\eta/s$ ratio extracted from BAMPS 
calculations with $k=0.5$, $1$, $3$, and $6$ via Eq. (\ref{eta}). 
$\eta/s$ decreases with increasing $k$. Each ratio is in good agreement
with a constant. These constant $\eta/s$ values and corresponding
$\sigma_{23}$ parametrizations are used for solving the hydrodynamic
equations (\ref{pi_2ndO}) (IS) and (\ref{pi_3rdO}) (third-order)
together with Eqs. (\ref{e}) and (\ref{n}).

\begin{figure}[ht]
\begin{center}
 \includegraphics[width=11cm]{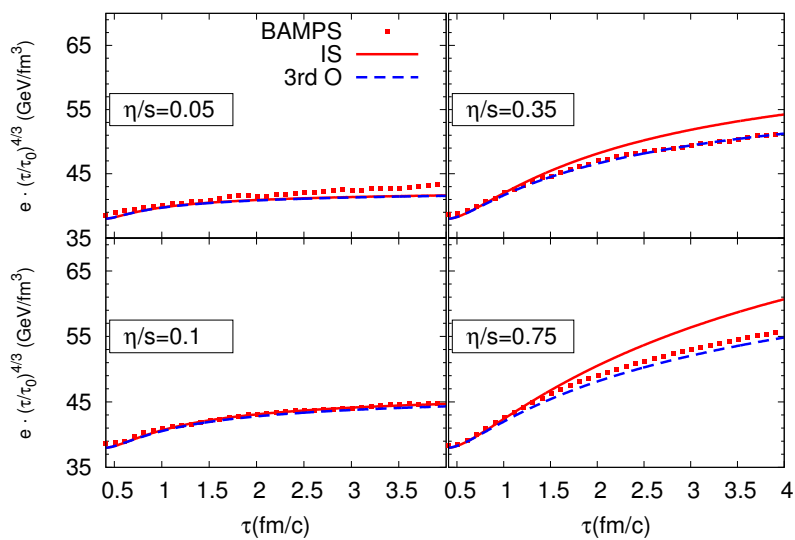}
\end{center}
\vspace{-0.5cm}
\caption{(Color online) Rescaled energy density from BAMPS, IS and
third-order hydrodynamic calculations.}
\label{fig:e}
\end{figure}

\begin{figure}[ht]
\begin{center}
 \includegraphics[width=11cm]{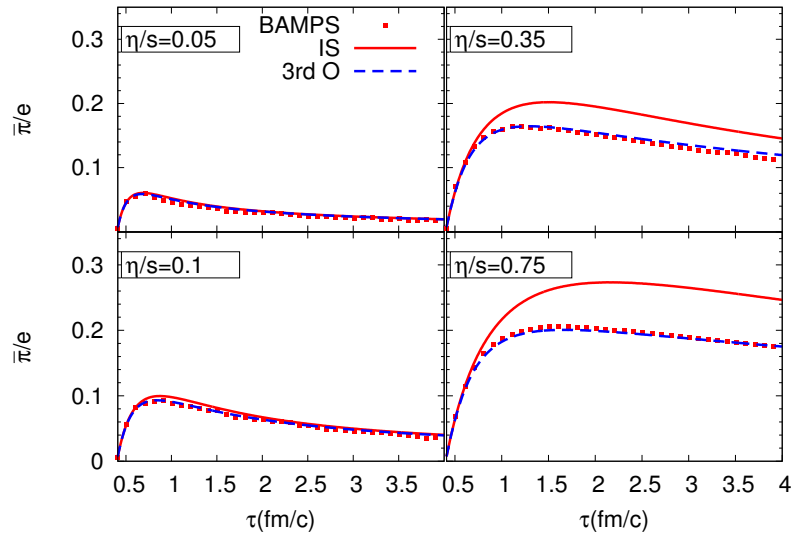}
\end{center}
\vspace{-0.5cm}
\caption{(Color online) Shear pressure to energy density ratio from BAMPS,
IS and third-order hydrodynamic calculations.}
\label{fig:Re}
\end{figure}

\begin{figure}[ht]
\begin{center}
 \includegraphics[width=11cm]{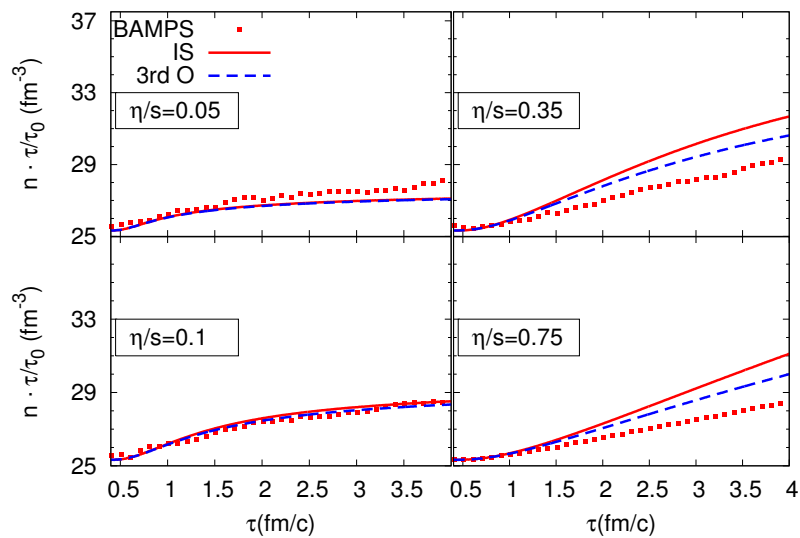}
\end{center}
\vspace{-0.5cm}
\caption{(Color online) Rescaled particle number density from BAMPS, IS
and third-order hydrodynamic calculations.}
\label{fig:dNdeta}
\end{figure}

\begin{figure}[ht]
\begin{center}
 \includegraphics[width=11cm]{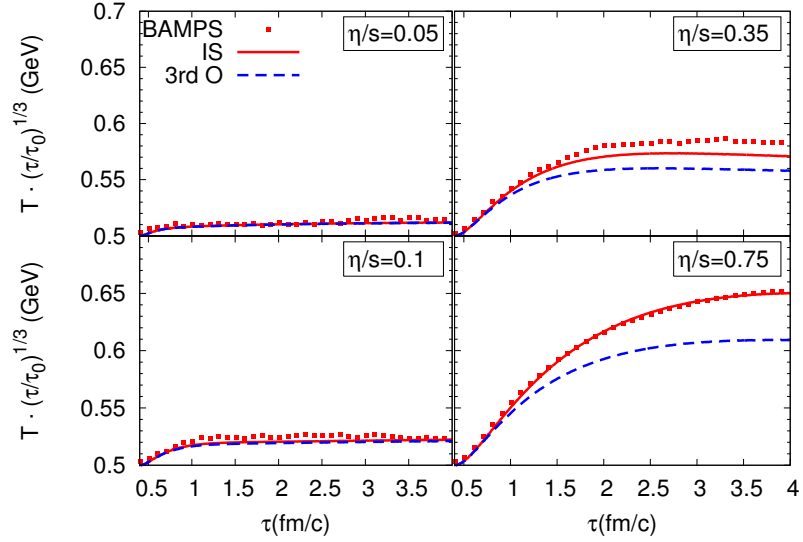}
\end{center}
\vspace{-0.5cm}
\caption{(Color online) Rescaled effective temperature $T=e/(3n)$ from BAMPS,
IS and third-order hydrodynamic calculations.}
\label{fig:T}
\end{figure}

\begin{figure}[ht]
\begin{center}
 \includegraphics[width=11cm]{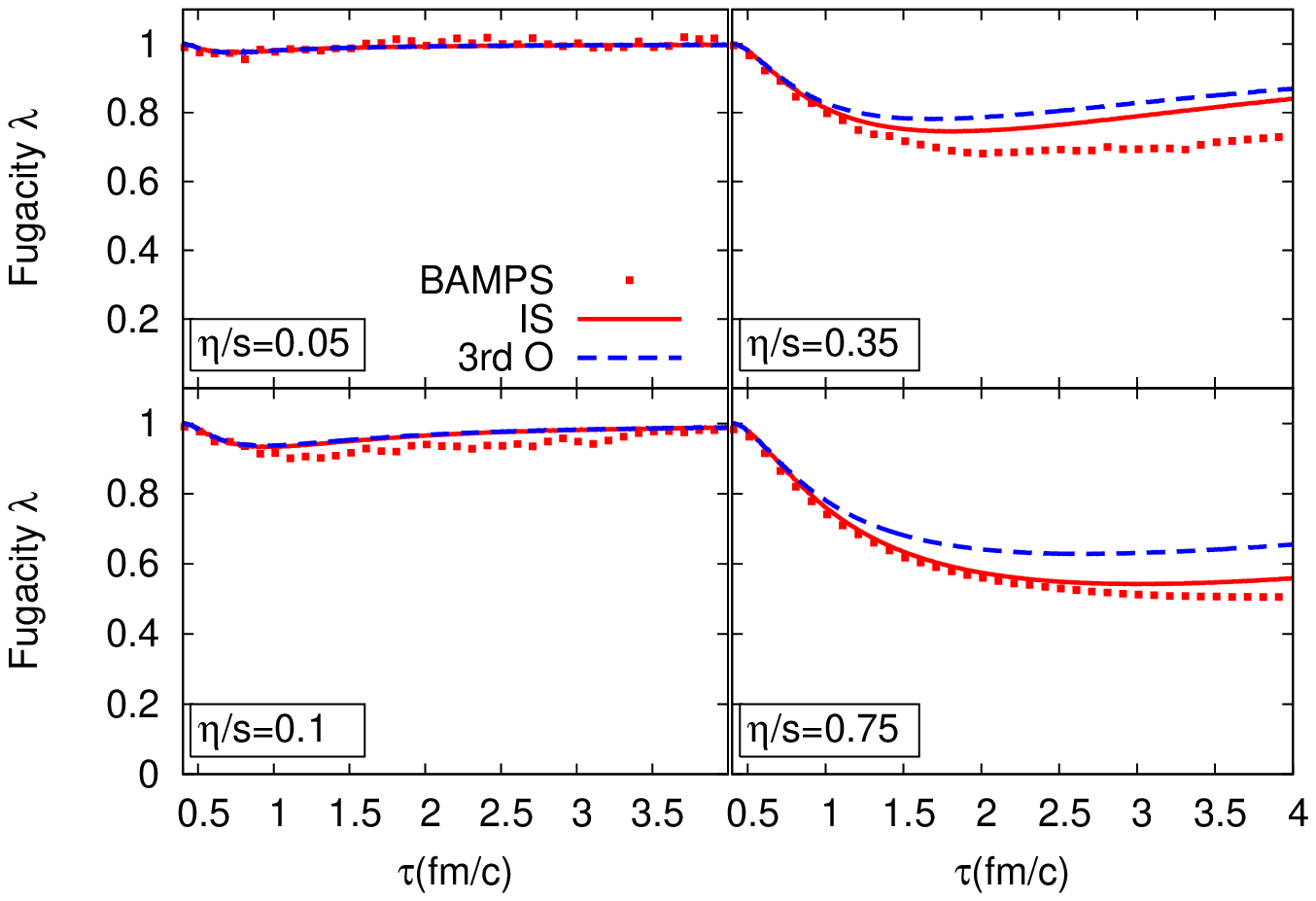}
\end{center}
\vspace{-0.5cm}
\caption{(Color online) Fugacity from BAMPS, IS and third-order hydrodynamic
calculations.}
\label{fig:fuga}
\end{figure}

Figures \ref{fig:e}, \ref{fig:Re}, \ref{fig:dNdeta}, \ref{fig:T},
and \ref{fig:fuga} show the comparisons between the hydrodynamic and
transport calculations for a one-dimensional boost-invariant expansion
of a gluon system with the thermal initial condition $T(\tau_0)=0.5$ GeV,
$\lambda(\tau_0)=1$ at $\tau_0=0.4$ fm/c. The results are rescaled in
the same manner as previously done in Fig. \ref{fig:comp}.

In Figs. \ref{fig:e} and \ref{fig:Re} we observe that the energy density
and the shear pressure from the third-order hydrodynamic calculations agree
perfectly with the BAMPS results even for large $\eta/s$ values. 
In contrast, the viscous effect is overestimated in the IS equations.
This becomes significant for large $\eta/s$ values,
which mark the applicability boundary of the IS approach. These findings
are in line with those in our previous work \cite{3rdO}, where particle
changing processes were not taken into account.

The difference in the particle number density (Fig. \ref{fig:dNdeta}) between 
the hydrodynamic and BAMPS results is large at large $\eta/s$ values.
Although both IS and third-order equations give larger densities than
those in BAMPS, the third-order results are closer to the ones from BAMPS .
On the other hand, looking at the temperature $T$ (Fig. \ref{fig:T})
and the fugacity $\lambda$ (Fig. \ref{fig:fuga}) we observe that the IS 
results show better agreement with the BAMPS results than the third-order
ones. However, it is difficult to make conclusions about applicability of
a hydrodynamic approach basing on the observables $T$ and $\lambda$. 
These quantities are defined via $e$ and $n$, but not solved directly 
from the hydrodynamic equations. If we consider the particle number 
conservation, the results on $T$ and $\lambda$ from the third-order
calculations are in very good agreement with those from BAMPS,
as can be cocluded from our observations in Ref. \cite{3rdO}. In the situation
considered here, small differences between hydrodynamic and kinetic
transport results on both key observables $e$ and $n$ translate into 
differences in $T$ and $\lambda$ in a non-trivial way.

\begin{figure}[ht]
\begin{center}
 \includegraphics[width=12cm]{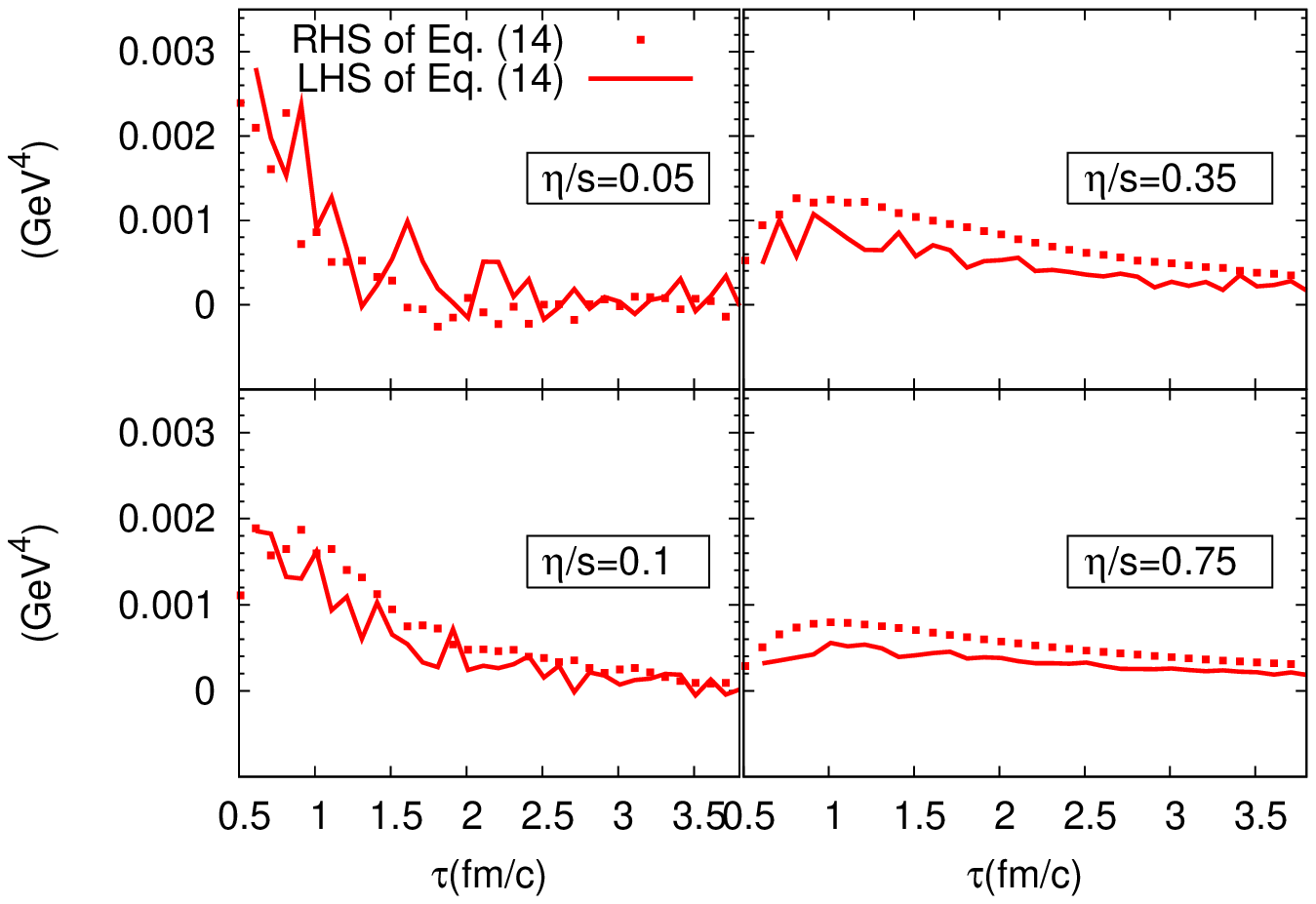}
\end{center}
\vspace{-0.5cm}
\caption{(Color online) Left and right hand sides of Eq.(\ref{approx})
(denoted by LHS and RHS, respectively)
calculated using the actual values of $R_{23}$, $R_{32}$, $n$, $\lambda$ and
$\sigma_{23}$ from BAMPS.}
\label{fig:rates}
\end{figure}

\begin{figure}[ht]
\begin{center}
 \includegraphics[width=12cm]{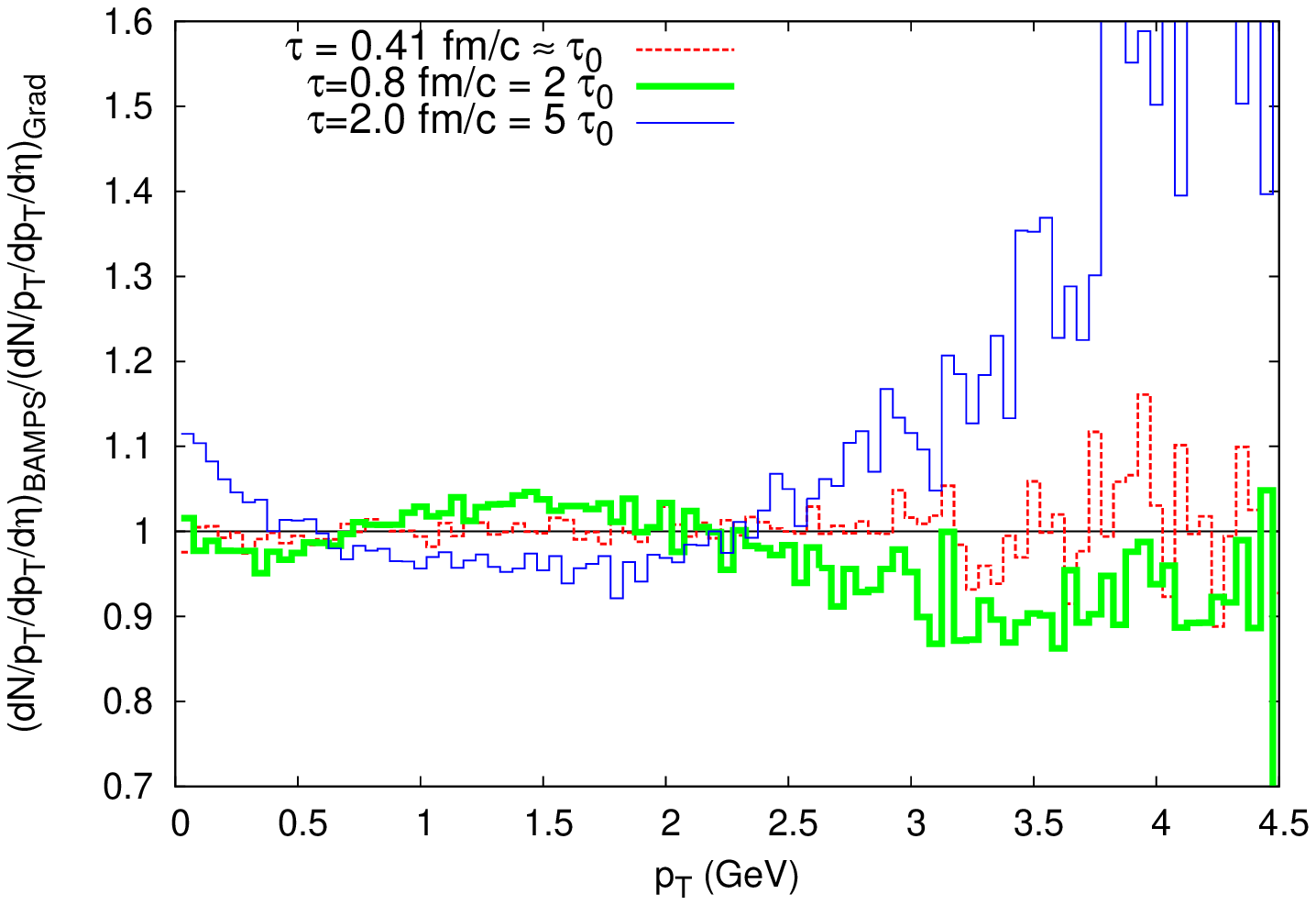}
\end{center}
\caption{(Color online) Ratio of transverse momentum spectrum extracted
in BAMPS to the one calculated using the Grad's approximation at different
 times with $\eta/s=0.35$.}
\label{fig:spectra_f1}
\end{figure}

To understand the differences in particle number densities between 
the hydrodynamic and BAMPS results, we examine Eq. (\ref{approx}), which
is valid by virtue of the approximation (\ref{f_offeq_1d}). For this we
calculate the left and the right hand sides of Eq. (\ref{approx}) using
the actual values of $R_{23}$, $R_{32}$, $n$, $\lambda$, and
$\sigma_{23}$ extracted from the BAMPS calculations. The results are shown 
in Fig. \ref{fig:rates}. Except for the case of $\eta/s=0.05$, 
$n^2 (1-\lambda)\sigma_{23}/2$ is always larger than $nR_{23}/2-nR_{32}/3$,
which leads to a stronger particle production in the hydrodynamic than
in the transport approach, as seen in Fig. \ref{fig:dNdeta}. Therefore,
it is obvious that the approximate distribution (\ref{f_offeq_1d}) must
deviate from the one extracted from BAMPS. This deviation is
demonstrated in Fig. \ref{fig:spectra_f1} by the ratio of $p_T$ spectra
at midrapidity extracted from BAMPS to the ones calculated using the
Grad's approximation (\ref{f_offeq_1d}). The results are
presented for $\eta/s=0.35$ at various times. The deviations of
the ratios from $1$ are on a level of $10\%-20\%$ and become more
pronounced at later times, $\tau\sim 1-2$ fm/c, at which the
viscous effect is strongest, see Fig. \ref{fig:Re}. We thus need
a modification of the Grad's approximation, which is a highly interesting
project for future studies.

\section{Summary and Outlook}
\label{conclusions}
In this work we have presented an extended set of viscous hydrodynamic 
equations, in which particle number changing processes are taken into 
account via a rate equation. The rate equation of the particle number 
density is derived from the Boltzmann equation employing the Grad's 
approximation. A one-dimensional boost-invariant expansion is considered
for simplifying the derivation of the hydrodynamic equations.
We have demonstrated that a proper treatment of particle number evolution
is essential for a hydrodynamic description of the gluon system, especially
for describing chemical equilibration and for determining the temperature
and the momentum spectra of the system. 

We have compared the results between the hydrodynamic and transport 
calculations including inelastic $2\leftrightarrow 3$ processes.
The energy density and the shear pressure obtained
from the third-order hydrodynamic equations agree well with the results
from the transport approach using BAMPS even at large $\eta/s=0.75$, while 
the results from the Israel-Stewart hydrodynamics deviate from the BAMPS
results by $10\%-20\%$ for $\eta/s=0.35-0.75$. 
Both the IS and the third-order hydrodynamic
calculations fail to meet the BAMPS results on the particle number density
at large $\eta/s$. The reason is that at large $\eta/s$ the viscous
effect is so large that the Grad's ansatz for the off-equilibrium
distribution should be modified beyond the second order in momentum,
in order to give a more accurate rate equation of the
particle number density. This is a highly interesting and timely subject
for future investigations.

The parton cascade BAMPS is presently being extended by including
light quark degrees of freedom \cite{Oliver}. Light quarks are expected
to equilibrate both kinetically and chemically slower than a pure gluon
system. Kinetic and chemical off-equilibrium effects will have significant
influence on the jet and bulk medium physics. They will also be highly
relevant concerning the slopes of transverse spectra and the yields of
dileptons and photons, which were measured at RHIC \cite{Adler:2005ig}
to extract the initial temperature for hydrodynamic approaches.

\section*{Acknowledgements}
 A.~E. gratefully acknowledges fellowship by the Helmholtz foundation 
and thanks the University of Cape Town Physics Department for its hospitality
during the stay in Cape Town, where this work was partly completed. A.~M. acknowledges the support
from the National Research Foundation through the NRF-DFG grant. The BAMPS simulations were
performed at the Center for Scientific Computing of Goethe University. This work is
supported by BMBF and by the Helmholtz International Center
for FAIR within the framework of the LOEWE program (LandesOffensive zur
Entwicklung Wissenschaftlich-\"okonomischer Exzellenz) launched
by the State of Hesse.


\begin{thebibliography}{99}

\bibitem{Voloshin:2008dg}
  S.~A.~Voloshin, A.~M.~Poskanzer and R.~Snellings,
  arXiv:0809.2949 [nucl-ex].

\bibitem{Romatschke:2007mq}
  P.~Romatschke and U.~Romatschke,
  Phys.\ Rev.\ Lett.\  {\bf 99}, 172301 (2007)
  [arXiv:0706.1522 [nucl-th]].

\bibitem{Song:2007ux}
  H.~Song and U.~W.~Heinz,
  Phys.\ Rev.\  C {\bf 77}, 064901 (2008)
  [arXiv:0712.3715 [nucl-th]].

\bibitem{Luzum:2008cw}
  M.~Luzum and P.~Romatschke,
  Phys.\ Rev.\  C {\bf 78}, 034915 (2008)
  [Erratum-ibid.\  C {\bf 79}, 039903 (2009)]
  [arXiv:0804.4015 [nucl-th]].

\bibitem{Heinz:2009xj}
  U.~W.~Heinz,
  arXiv:0901.4355 [nucl-th].

\bibitem{Teaney:2009qa}
  D.~A.~Teaney,
  arXiv:0905.2433 [nucl-th].

\bibitem{Molnar:2008fv}
  E.~Molnar,
  Eur.\ Phys.\ J.\  C {\bf 60}, 413 (2009)
  [arXiv:0807.0544 [nucl-th]].

\bibitem{Molnar:2009tx}
  E.~Molnar, H.~Niemi and D.~H.~Rischke,
  Eur.\ Phys.\ J.\  C {\bf 65}, 615 (2010)
  [arXiv:0907.2583 [nucl-th]].

\bibitem{HM09}
P. Huovinen, D. Molnar, Phys.\ Rev.\ C\ 79 (2009) 014906.

\bibitem{Biro93}
T. S. Biro et al, Phys.\ Rev.\ C\ 48 (1993) 1275.

\bibitem{Srivastava:1996qd}
  D.~K.~Srivastava, M.~G.~Mustafa and B.~M\"uller,
  Phys.\ Rev.\  C {\bf 56}, 1064 (1997)
  [arXiv:nucl-th/9611041].

\bibitem{ER2000}
D. M. Elliott, D. H. Rischke, Nucl.\ Phys.\ A\ 671 (2000) 583 

\bibitem{Xu:2004mz}
  Z.~Xu and C.~Greiner,
  Phys.\ Rev.\  C {\bf 71}, 064901 (2005)
  [arXiv:hep-ph/0406278].

\bibitem{Gelis:2005pb}
  F.~Gelis, K.~Kajantie and T.~Lappi,
  Phys.\ Rev.\ Lett.\  {\bf 96}, 032304 (2006)
  [arXiv:hep-ph/0508229].

\bibitem{Dusling:2008xj}
  K.~Dusling and S.~Lin,
  Nucl.\ Phys.\  A {\bf 809}, 246 (2008)
  [arXiv:0803.1262 [nucl-th]].

\bibitem{Martinez:2008di}
  M.~Martinez and M.~Strickland,
  Phys.\ Rev.\  C {\bf 78}, 034917 (2008)
  [arXiv:0805.4552 [hep-ph]].

\bibitem{vanHees:2009vk}
  H.~van Hees and R.~Rapp,
  Nucl.\ Phys.\  A {\bf 827}, 341C (2009)
  [arXiv:0901.2316 [nucl-th]].

\bibitem{Schenke:2006yp}
  B.~Schenke and M.~Strickland,
  Phys.\ Rev.\  D {\bf 76}, 025023 (2007)
  [arXiv:hep-ph/0611332].

\bibitem{Lin:2002rw}
  Z.~w.~Lin and C.~M.~Ko,
  Phys.\ Rev.\ Lett.\  {\bf 89}, 202302 (2002)
  [arXiv:nucl-th/0207014];
  V.~Greco, C.~M.~Ko and P.~Levai,
  Phys.\ Rev.\ Lett.\  {\bf 90}, 202302 (2003)
  [arXiv:nucl-th/0301093];
  Phys.\ Rev.\  C {\bf 68}, 034904 (2003)
  [arXiv:nucl-th/0305024]; 
  R.~J.~Fries, B.~M\"uller, C.~Nonaka and S.~A.~Bass,
  Phys.\ Rev.\ Lett.\  {\bf 90}, 202303 (2003)
  [arXiv:nucl-th/0301087];
  Phys.\ Rev.\  C {\bf 68}, 044902 (2003)
  [arXiv:nucl-th/0306027];
  D.~Molnar and S.~A.~Voloshin,
  Phys.\ Rev.\ Lett.\  {\bf 91}, 092301 (2003)
  [arXiv:nucl-th/0302014];
  R.~C.~Hwa and C.~B.~Yang,
  Phys.\ Rev.\  C {\bf 67}, 064902 (2003)
  [arXiv:nucl-th/0302006].

\bibitem{Dutta:1999cn}
  D.~Dutta, A.~K.~Mohanty, K.~Kumar and R.~K.~Choudhury,
  Phys.\ Rev.\  C {\bf 61}, 034902 (2000)
  [arXiv:hep-ph/9908359].

\bibitem{El09}
  A.~El, A.~Muronga, Z.~Xu and C.~Greiner,
  Phys.\ Rev.\  C {\bf 79}, 044914 (2009)
  [arXiv:0812.2762 [hep-ph]].

\bibitem{3rdO}
  A.~El, Z.~Xu and C.~Greiner,
  Phys.\ Rev.\  C {\bf 81}, 041901 (2010)
  [arXiv:0907.4500 [hep-ph]].

\bibitem{Monnai:2010qp}
  A.~Monnai and T.~Hirano,
  arXiv:1003.3087 [nucl-th].

\bibitem{Bouras:2009nn}
  I.~Bouras {\it et al.},
  Phys.\ Rev.\ Lett.\  {\bf 103}, 032301 (2009)
  [arXiv:0902.1927 [hep-ph]];
  Nucl.\ Phys.\  A {\bf 830}, 741C (2009)
  [arXiv:0907.4519 [hep-ph]];
  arXiv:1006.0387 [hep-ph];
  I.~Bouras {\it et al.},
  Acta Phys.\ Polon.\  B {\bf 40}, 973 (2009).

\bibitem{Denicol:2010xn}
  G.~S.~Denicol, T.~Koide and D.~H.~Rischke,
  arXiv:1004.5013 [nucl-th].

\bibitem{IS}
W.~Israel, Ann.\ Phys.\ (N.Y.)  100 (1976) 310;
J.M.~Stewart, Proc.\ R.\ Soc.\ London\ , Ser.\ A\  357 (1977) 59;
W.~Israel, M.~Stewart, Ann.\ Phys.\ (N.Y.) 118 (1979) 341.

\bibitem{M04}
A. Muronga, Phys.\ Rev.\ C\ 69 (2004) 034903.

\bibitem{M07}
A. Muronga, Phys.\ Rev.\ C 76 (2007) 014910.
%

\bibitem{Muronga05}
A. Muronga, J.\ Phys.\ G\ 31 (2005) 1035-1039.

\bibitem{Xu:2007aa}
  Z.~Xu and C.~Greiner,
  Phys.\ Rev.\  C {\bf 76}, 024911 (2007)
  [arXiv:hep-ph/0703233].

\bibitem{Xu:2010cq}
  Z.~Xu and C.~Greiner,
  Phys.\ Rev.\  C {\bf 81}, 054901 (2010)
  [arXiv:1001.2912 [hep-ph]].

\bibitem{XuPRL08}
Z. Xu, C. Greiner, Phys.\ Rev.\ Lett.\ 100 (2008) 172301.

\bibitem{Oliver}
O.~Fochler, Z.~Xu, C.~Greiner, in preparation. 

\bibitem{Adler:2005ig}
  S.~S.~Adler {\it et al.}  [PHENIX Collaboration],
  Phys.\ Rev.\ Lett.\  {\bf 94}, 232301 (2005)
  [arXiv:nucl-ex/0503003];
  A.~Adare {\it et al.}  [PHENIX Collaboration],
  Phys.\ Rev.\  C {\bf 81}, 034911 (2010)
  [arXiv:0912.0244 [nucl-ex]].


\end{thebibliography}
\end{document}